\documentclass[pdfmx]{aastex631}

\usepackage{amsmath, xparse}
\usepackage{tikz}

\shorttitle{Fractal algorithm for multiple lens analyses.}
\shortauthors{F. Abe}
\graphicspath{{./}{figures/}}

\begin{document}

\title{Fractal algorithm for multiple lens analyses
}

\author[0009-0000-8865-4778]{F. Abe}
\affiliation{Institute for Space-Earth Environmental Research, ISEE, Nagoya University \\
Furo-cho, Chikusa-ku, Nagoya-shi \\
Aichi, 464-8601, Japan}
\email {abe@isee.nagoya-u.ac.jp, fabe.earth@gmail.com}

\begin{abstract}

A microlensing exoplanet search is a unique method for finding planets orbiting distant stars.
However, in the past, the method used to analyze microlensing data could not deal with complex lens systems. 
The number of lenses was limited to three or less. 
Positions calculations of images and their integration suffered from severe round-off errors 
because of singularities. 
We developed a new algorithm to calculate the light curves of multiple lens systems.
In this algorithm, fractal-like consecutive self-similar division (SSD) is used to find sparse images. 
SSD is also useful for integrating images to efficiently obtain magnifications.
The new algorithm does not use root finding for the lens equation and  is free from caustic singularities. 
There is no limit on the number of lenses that can be used with this algorithm. 
Compared to inverse-ray shooting, this  method dramatically improves the computing time.
The calculation can be adjusted to obtain either a high-precision final result or  high-speed quick result. 
Although this new algorithm was developed for a microlensing planet search, its application to quasar microlensing 
is also expected. 
This paper discusses problems in the modeling of a multiple lens system and then presents the new algorithm in detail.

\end{abstract}

\keywords{Gravitational lenses --- Extra solar planets --- Quasars --- Numerical method and analysis}

\section{Introduction} \label{sec:intro}

The microlensing planet search method was introduced by \citet{1991ApJ...374L..37M}. 
Gravitational microlensing \citep{1936Sci....84..506E,1986ApJ...304....1P} 
is the magnification of the apparent brightness of a 
distant star (source star) caused by the gravitational lensing of another star (lensing object) passing 
between the source star and the observer. 
If the lensing object has planets, the perturbations 
caused by these planets may be observed. The configuration of the planets and 
host star is obtained by a detailed analysis of the light curve. 
A collaboration between the microlensing observations in astrophysics (MOA) project 
and optical gravitational lensing experiment (OGLE) discovered the first planet 
\citep{2004ApJ...606L.155B} using this method. 
Since then, more than 100 planets have been discovered using this method. 
This method is particularly sensitive \citep{1992ApJ...396..104G,1996ApJ...472..660B} 
to a planet projected near the Einstein 
ring radius of the host star. 
In this region, microlensing is even sensitive to low mass planets 
down to earth mass or less. 
For typical lens-source geometries, this region is 
right outside of the snow line and expected to be  
the birth place \citep{2004ApJ...604..388I} of giant planets. 
Thus, the microlensing planet search is expected to be useful in investigating the formation process 
of outer planets.
In the near future, space telescopes like the Nancy Grace Roman 
telescope (\url{https://roman.gsfc.nasa.gov/}) 
will be used for 
microlensing planet searches. 
Because of the improved precision of photometry 
and cadence of observations, 
the discovery of numerous planets, 
including multiple planet systems and  
exomoons (extra-solar giant natural satellites 
\citep{2002ApJ...580..490H,2010A&A...520A..68L,2022A&A...664A.136B,
2019ApJS..241....3P}  
is expected. 
However, high-precision photometry and high-cadence observations 
generate new challenges for analyses, 
increasing the importance of high-precision and high-speed 
modeling for multiple lens analyses \citep{2025A&A...694A.219B}. 

Modeling a microlensing event is a well-defined geometrical problem but is challenging. 
Figure \ref{fig:configuration} shows a configuration of a multiple lens 
system. 
Here we assume that all of the lenses 
are point-like lying on the same plane (lens plane). 
A point on the lens plane and corresponding point on the source plane are related by the lens equation.
This equation for a multiple point lens system is written as follows:

\begin{equation} \label{eq:lens}
          \vec{\beta} = \vec{\theta} - \sum^{n}_{i=1} q_i \frac{\vec{\theta} - \vec{l_i}}{|\vec{\theta} - \vec{l_i}|^2},
\end{equation}
where, $\vec{l_i}$ (i = 1, .., n) are the position vectors of the ensemble 
of $n$ lenses, $\vec{\theta}$ is a position vector on the 
lens plane, $\vec{\beta}$ is the mapped position vector on the source plane, and $q_i$ is 
the mass ratio of the ith lens to the total mass.
All of the vectors are normalized by the angular Einstein radius:

\begin{figure}[ht!]
\plotone{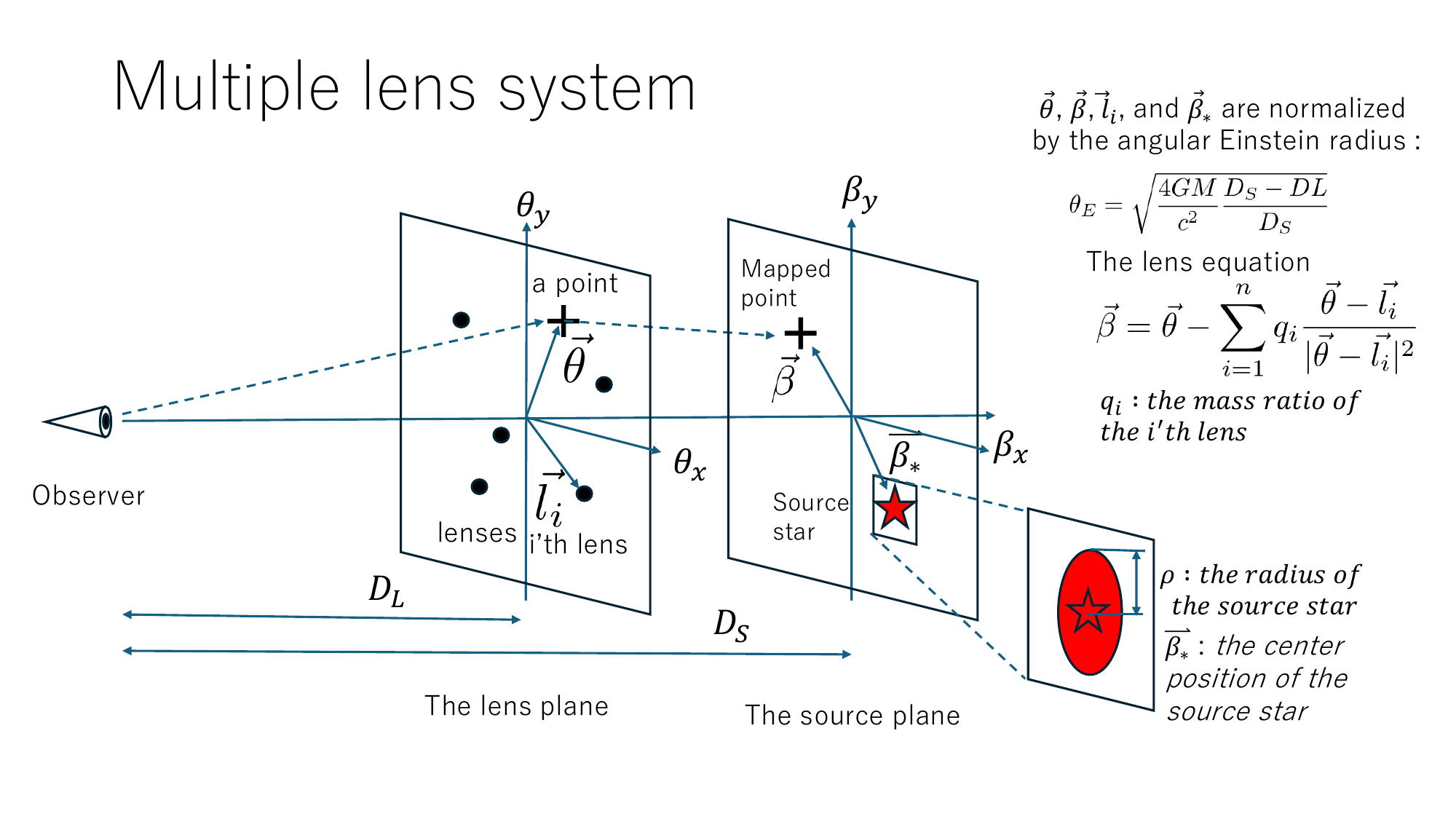}
          \caption{Configuration of the multiple lens system. 
          All of the lenses are assumed to lie within  
          a plane that is orthogonal to the line of sight (lens plane).
 \label{fig:configuration}}
\end{figure}

\begin{equation} \label{eq:thetaE}
          \theta_E = \sqrt{\frac{4GM}{c^2}\frac{D_S - D_L}{D_S}},
\end{equation}
where, $G$ is the gravitational constant, $M$ is the total mass of the lens system, $c$ is the speed of light, 
$D_S$ 
is the distance between the observer and source, and $D_L$ is the distance between the observer and the lens.
Although the magnification produced by the lens system is often calculated based on 
the conservation of the surface brightness (Liouville's Theorem), 
this calculation is not straightforward. 
A point on the source plane corresponds to multiple points on the lens plane. 
To find image positions, root finding for the lens equation is necessary. 
However root finding is difficult for a large number of lenses and sometimes causes severe round-off errors 
\citep{2025A&A...694A.219B} 
because of singularities.

During real microlensing events, the source star has a finite size. 
The shape is circular or slightly elliptical. 
The images produced by the lens system are deformed and have crescent shapes. 
Because the surface brightness is conserved during the gravitational lensing, 
the ratio of the total area of the images 
to the area of the source star corresponds to the magnification for a uniform surface brightness. 
In a typical bulge microlensing event, the angular radius of the source star is approximately 
one $\mu$as. 
That is much smaller than the typical angular Einstein radius, which is approximately 
several hundred $\mu$as.
Thus, the produced images are very sparse and small compared with the angular Einstein radius. 
Point source approximation has been thought to be effective in obtaining a rough result to begin an analysis. 
The magnifications of the images for a point source are calculated using the Jacobian.
The Jacobian determinant, $J(\vec{\theta})$, is written as follows:
\begin{equation} \label{eq:J}
          \begin{split}
                    J(\vec{\theta}) &= \begin{vmatrix} 
          \frac{\partial \beta_x}{\partial \theta_x} & \frac{\partial \beta_x}{\partial \theta_y} \\
          \frac{\partial \beta_y}{\partial \theta_x} & \frac{\partial \beta_y}{\partial \theta_y} 
          \end{vmatrix} \\
          &= \biggl[1-\sum^{n}_{i=1}q_i\frac{(\theta_y - l_{i,y})^2 - (\theta_x - l_{i,x})^2}{|\vec{\theta} - \vec{l_i}|^4} \biggl]
\biggl[1-\sum^{n}_{i=1}q_i\frac{(\theta_x - l_{i,x})^2 - (\theta_y - l_{i,y})^2}{|\vec{\theta} - \vec{l_i}|^4} \biggl] \\
          &- \biggl[\sum^{n}_{i=1}2q_i\frac{(\theta_x - l_{i,x}) (\theta_y - l_{i,y})}{|\vec{\theta} - \vec{l_i}|^4}\biggl]^2.
          \end{split} 
\end{equation}
The total magnification $A$ is found as follows:
\begin{equation} \label{eq:A}
          A = \sum^{m}_{j=1} \frac{1}{|J(\vec{\theta_j})|},
\end{equation}
where $\vec{\theta_j}$ is the position of the jth image, and m is the number of images, 
$n+1 \ge m \ge 5n +1$ \citep{2003astro.ph..5166R}, 
depending on the configuration of the lenses.
In the region for which $J(\vec{\theta}) > 0$, the gravitational lensing produces erect images. 
On the other hand, in the region for which $J(\vec{\theta}) < 0$, the images are inverted. 
Critical curves are defined 
as the borders of the erect and inverted images where $J(\vec{\theta}) = 0$. 
Around the critical curves,  
the magnification become very high.
The critical curves mapped to the source plane are called caustics. 
As shown in the Eq.\ref{eq:A}, magnification $A$ diverges to infinity when a point source is on a caustic.
This singularity can cause severe problems in the analysis. 
Near a caustic, a small difference on the source 
plane induces a large difference on the lens plane. 
A small error on the source plane is magnified to become a large error if an attempt is made to solve 
the lens equation to obtain $\vec{\theta}$. 
Thus, calculations with very high precision have been necessary in previous analyses. 
In the case of high-magnification or sharp caustic crossing events, the finite size of 
the source star becomes critical. 
Previous algorithms adopted a step-by-step approach. 
Initially, the lens equation is solved for $\vec{\theta}$ to obtain 
the positions of the images. 
Then, integration's are performed to obtain the total magnification. 
However, the root finding of the lens equation is difficult. 
For a single lens ($n = 1$), a simple second order polynomial equation 
\citep{1986ApJ...304....1P} is derived from the lens equation.  
For a binary lens event, a fifth order polynomial equation \citep{1995ApJ...447L.105W} is obtained. 
\citet{2002astro.ph..2294R} obtained a tenth order polynomial equation for 
a triple lens system.  
However, similar algebraic method have not been discovered for more complex lens systems. 

Integrating the images is another difficult problem. \citet{2010ApJ...716.1408B} introduced 
image-centered inverse-ray shooting. 
In this method numerous uniform rays are produced on the lens plane 
and traced to the source plane to integrate the images.  
As another method, Stokes integration was proposed by \citet{1997ApJ...477..580G}.  
In this method, the boundary points are calculated as roots of the lens equation, and the integration 
uses  the Stokes theorem. 
However, image-centered inverse-ray-shooting requires significant computation, 
and a non-uniform 
surface brightness particularly stellar spots is hard to handle during the Stokes integration. 
These calculations must be repeated for different source and/or lens positions 
to construct light curves. 
Numerous light curves for various parameters are necessary to compare the results of observations. 
To analyze a planetary event, the optimum parameters for an extremely large parameter space are necessary. 
This analysis is very time consuming. 
Increasing the speed of the integration is critical for a fast analysis. 

Another approach to multiple lens analyses is inverse-ray shooting 
(IRS, \citet{1986A&A...164..237S}, \citet{1986A&A...166...36K}). 
In this method, numerous uniform light rays are produced on the lens plane 
and traced back to the source plane using the lens equation. 
The density of the rays on the source plane represents the magnification. 
Because this method does not use a root-finding approach, 
there are several advantages compared to conventional methods. 
There is no limitation on the number of lenses. The finite source effect can easily be implemented, 
and there is no caustic singularity because this method does not need  
to solve the lens equation. 
However, this method needs 
a large amount of computing power for blind shooting 
numerous rays. 
Until now, IRS is used with map making method \citep{2006ApJ...642..842D} 
if the motions of the lenses can be ignored. 

This paper presents a new algorithm for multiple lens systems. 
This method can be regarded as significant modification of IRS and/or adaptive contouring 
(\citet{2007MNRAS.377.1679D}) 
and does not use the root-finding approach. 
Thus, there is no limitation on the number of lenses. Because this method does not 
need to solve the lens equation, 
there is no critical round-off error problem  
caused by caustic singularities. 
Unlike IRS, this method uses a feed-back process to reduce the total calculations. 
This feed-back can be implemented using a fractal-like self-similar division (SSD) process. 
The computing time is dramatically reduced using this algorithm.

\section{Principle} \label{sec:principle}

As discussed in section \ref{sec:intro}, the "no-root-finding" approach has advantages, 
including the absence of 
singularities and no limits on the number of lenses. 
Of course, there are singularities at the centers of the lenses ($\vec{l_i}$, i = 1, ..., n),
but these are not critical. The positions around the centers of the lenses are mapped far from the source. 
We can simply ignore these singularities. Then, we are free from singularities. 
The remaining problem is how to reduce the computing time. 

\begin{figure}[ht!]
\plotone{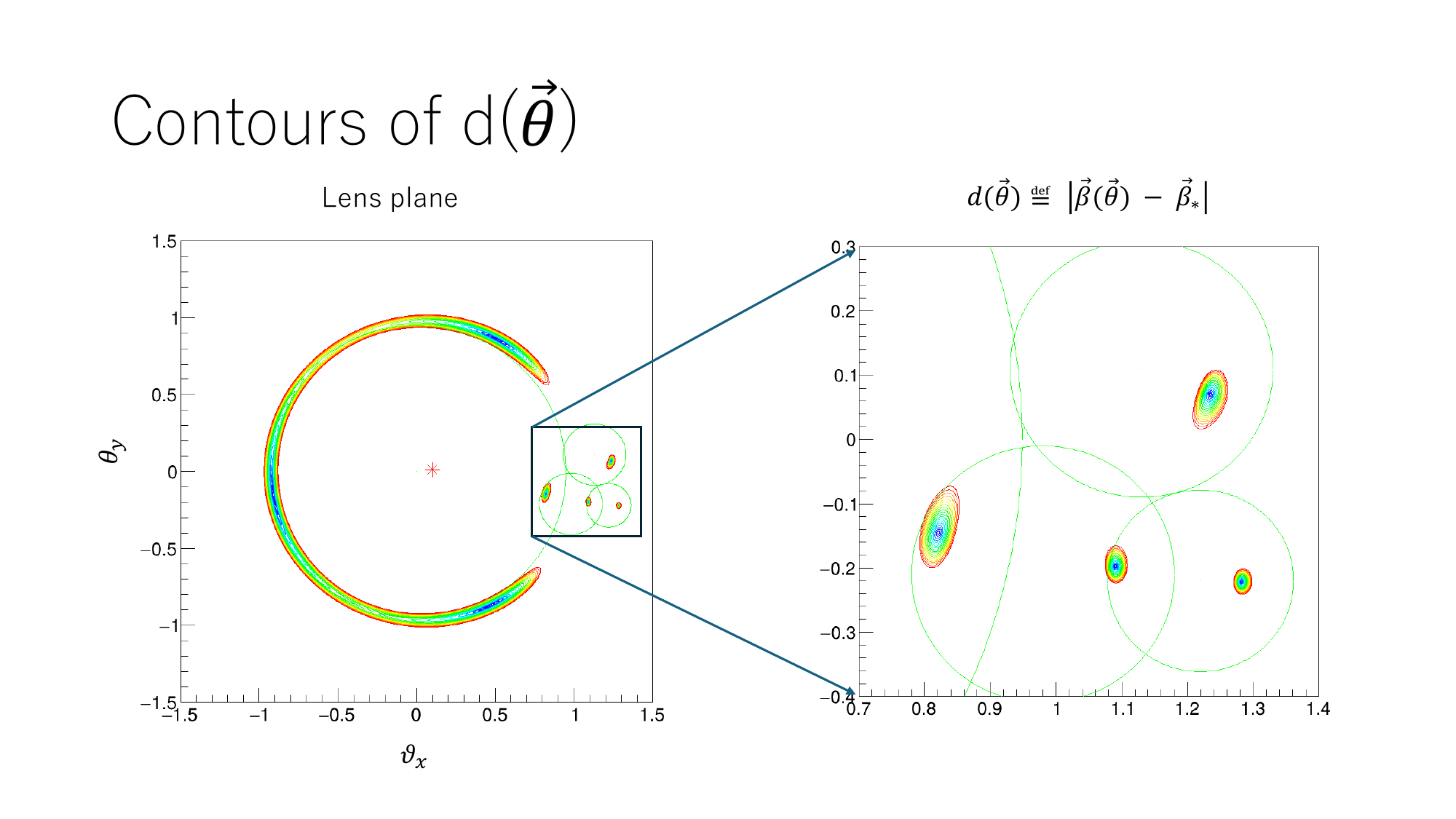}
          \caption{Contour map of $d(\vec{\theta}) = |\vec{\beta}(\vec{\theta}) - \vec{\beta_*}|$.
          The left panel shows the total area. The right panel shows the magnified view around 
          the small lenses. 
          The green circles show individual angular Einstein radii ($\theta_{E,i} = \sqrt{q_i}$). 
          The source position is shown as red asterisk. 
          The contours are truncated at $d(\vec{\theta}) = 0.1$ (shown by red) to emphasize small images. 
          Yellow, green, and blue contours show smaller values of $d(\vec{\theta})$.
 \label{fig:contour}}
\end{figure}

Let us consider a point on lens plane $\vec{\theta}$, the corresponding point on source plane 
$\vec{\beta}(\vec{\theta})$, and a source star centered at $\vec{\beta_*}$. 
The shape of the source star may be circular, or slightly elliptical. 
We assume here that we can clearly identify whether $\vec{\beta}(\vec{\theta})$ is inside or outside of the source star. 
If $\vec{\beta}(\vec{\theta})$ is inside of the source, we can identify $\vec{\theta}$ as inside of an image. 
If $\vec{\beta}(\vec{\theta})$ is outside of the source, we are not interested in it because it is outside of the images 
and so does not contribute to the total area of the images. 
If we can produce multiple examples of $\vec{\theta}$ and the corresponding examples of 
$\vec{\beta}(\vec{\theta})$, 
we can reproduce the shapes of 
the images by identifying those examples of $\vec{\beta}(\vec{\theta})$ that are 
inside or outside of the surface of the source. 
We define the distance between the mapped point and the center of the source as
\[
d(\vec{\theta}) = \left| \vec{\beta}(\vec{\theta}) - \vec{\beta}_* \right|.
\]
Figure~\ref{fig:contour} displays the contour map of \( d(\vec{\theta}) \).
The mass ratios and positions of the lenses are assumed to be
\[
(q_i, (l_{x,i}, l_{y,i})) = (0.90, (0.0, 0.0)),\ (0.04, (1.13, 0.11)),\ (0.04, (0.98, -0.21)),\ \text{and}\ (0.02, (1.22, -0.22)).
\]
The source position is set to \( \vec{\beta}_* = (0.1, 0.01) \).
To highlight regions with small values of \( d(\vec{\theta}) \), the contours are truncated at \( d(\vec{\theta}) = 0.1 \).
As shown in the figure, the images are clearly identified by the contours. 
The contours for $ d(\vec{\theta}) = \rho$ show 
the images for the source star, where $\rho$ is the angular radius of the source star 
divided by the angular Einstein radius. 

The problem is how to find those images quickly. 
Of course, a brute-force grid approach wastes time because most of the points are outside of the images. 
To reduce unproductive computations, we must remove the regions on the lens plane where 
there is no image, which can  
be done by identifying regions mapped far from the source.

To identify regions of interest or non-interest, we consider a small triangle on the lens plane with vertices at $\vec{\theta}_1$, $\vec{\theta}_2$, and $\vec{\theta}_3$. 
The corresponding points on the source plane are denoted as $\vec{\beta}_1$, $\vec{\beta}_2$, and $\vec{\beta}_3$, respectively, obtained via the lens mapping.
If all corners of the triangle lie securely within the source, the triangle is identified as being inside the source. 
Conversely, if all sides of the triangle lie securely outside the source and the center of the source is not within the triangle, it is identified as being outside the source.
Geometrically, a triangle can be regarded as being outside a circular source if all of its sides lie outside the circle 
and the center of the circle is not enclosed by the triangle. However, due to gravitational lensing, the mapping from the lens plane to the source plane may deform the triangle, and thus this simple geometric criterion may not hold.
Due to gravitational lensing, the mapped triangle may be deformed. Therefore, a proper margin is necessary to ensure reliable identification. 
The margin used to determine whether a triangle is inside or outside the source was set to 20\% of the triangle's side length.  
This means the margin is relatively large for large triangles and small for small ones.  
The validity of this choice can be verified through the integrity check described in Section~\ref{sec:algorithm}.
Because of the finite size of the triangle and the margin, it may be challenging to determine whether triangles near the source's boundary are inside or outside. 
This uncertainty is expected to be smaller for smaller triangles.
If a triangle is securely identified as outside, it can be discarded. 
If the triangle is securely inside, it is retained for calculating the total image area. 
If it cannot be determined whether the triangle is inside or outside, it suggests that the triangle lies on or near the boundary. 
In such cases, the triangle can be subdivided into smaller triangles to more precisely determine the boundary position. 
Repeating this process makes it possible to determine the shapes of images.

The method we propose here is to repeat fractal-like self-similar divisions to obtain the shapes of the images. 

\section{Algorithm} \label{sec:algorithm}
\begin{figure}[ht!]
\plotone{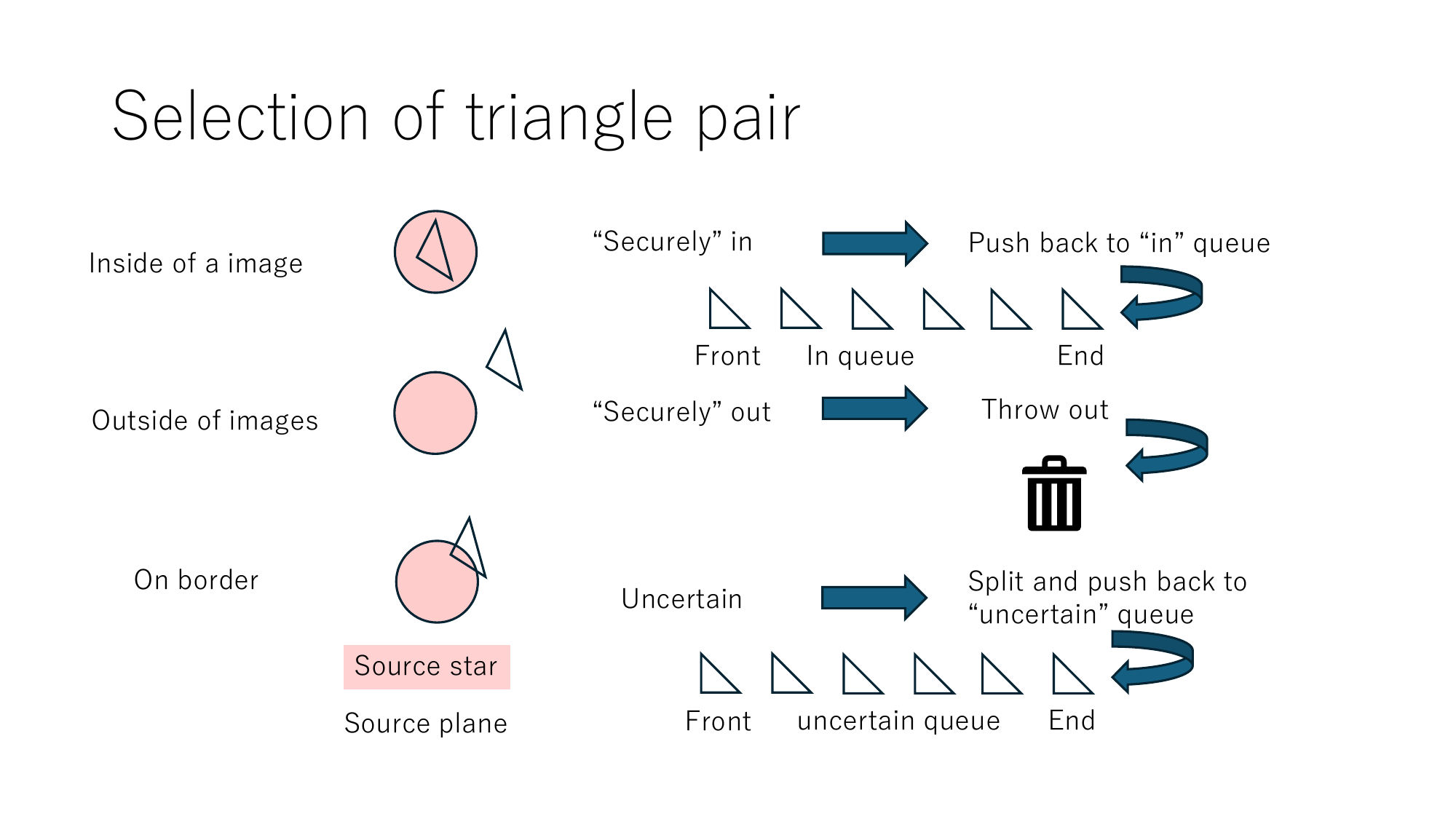}
          \caption{Schematic diagram of the selection process. 
          There are three queues of 
          triangle pairs (TriMaps): 
          in, out and uncertain. Ordinarily, we do not use those in the out queue. 
          Initially, all of the pairs are listed in the uncertain queue. The first element 
          in the uncertain queue is analyzed and identified as in, out, or uncertain. 
          If the element is in, it is pushed back to the end of the in queue. If the element is out, 
          it is thrown out. If the element is uncertain, it is divided into two TriMaps, 
          which are pushed back to 
          the end of the uncertain queue. 
 \label{fig:selection}}
\end{figure}

We invented an analysis algorithm for multiple lens system. In this algorithm, we define pairs of 
triangles (TriMaps), each of which consists of a triangle on the lens plane (TriL) 
and corresponding triangle on the source plane (TriS). 
Then we define three queues of TriMaps: in, out, and uncertain. 
Figure \ref{fig:selection} shows a schematic diagram of the process. 
The ``in'' queue is a collection of TriMaps whose TriSs are securely inside of the source star. 
The ``out''  queue is a collection of TriMaps whose TriSs are securely outside of the source star. 
The  ``uncertain'' queue is a collection of TriMaps whose TriSs are on the border 
of the source star or not securely identified. 
Ordinarily, we are not interested in the ``out'' queue. This queue is only used for the purpose 
of an integrity check. 
At the beginning of an analysis, we define a square area to calculate. 
This area must be large enough to cover all of the images. 
If the source star is close to or inside of the lens system, the images appear inside 
or  around the Einstein ring $\theta_E$. 
If the source star is far from the lens system, one of the images appears close to the source. 
However, the approximate image position can be obtained by a single lens approximation. 
Then we can define a square area to search the images. 
To start to search the images, we divide the square area into 2048 right angled 
isosceles triangles as the initial TriLs.  
Then corresponding TriSs are calculated. 
First, these initial triangle pairs are lined up in the  
``uncertain''  queue. 

Then we begin the search process to identify the images. The first triangle pair in the uncertain 
queue 
is analyzed 
to identify whether the TriS is 
securely in, securely out, or uncertain. 
If the pair is identified as securely in, the pair is moved to the ``in'' queue. 
If the pair is securely out, it is simply removed unless we attempt an integrity check. 
If the pair is identified as uncertain, the TriL is divided into two triangles 
with similar shapes (right angle isosceles). 
The new TriL point caused by this dividing process is mapped on 
the source plane to divide the TriS into two.
The new triangle pairs are lined up at the end of the uncertain queue. 
The same processing is done for all of the initially lined triangle pairs. 
After this process, all of the triangle pairs in the uncertain queue become half their original size. 
We call this consecutive processing a "round". 
We can repeat this process multiple times to reduce the area of a TriL. 
As the size of the TriLs decrease, some of the pairs will be 
identified as in or out. Thus the total TriL area in the uncertain queue is expected to decrease. 
If the size of the TriLs in the uncertain queue becomes sufficiently small, 
all of the images are expected to be picked up.

During these processes, there are three sources of confusion. Confusion is caused by 
a triangle that is "too large" in size on the lens
plane compared to the size of the lens. 
Just as areas can be inside and outside of the critical curve, 
the lens system produces images with "opposite" 
directions. 
Thus a large triangle around a critical curve may cause confusion. The critical curves 
occur close to the angular Einstein rings of the lenses ($\theta_{E,i} = \sqrt{q_i}$). 
To avoid this confusion, a large triangle that is close to a lens is 
identified as uncertain regardless of whether the corresponding triangle 
on the source plane is 
in, out, or uncertain. 
The second source of confusion appears in the center of a lens. 
If a triangle includes the lens center, the inside of the triangle 
is mapped to outside of the corresponding triangle on the source plane. 
Thus if a small triangle includes the center 
of the lens and the corresponding triangle on the source plane includes the source star, 
this pair must be 
identified as  ``out''. 
This processing may result in the loss of very small images close to the center of a lens. 
However, the contribution to the total magnification is expected to be very small 
because those images are very small. 
The third source of confusion appears when the source is far from the lens system. In this case, a small image appears outside 
of the source. To pick up this small image, we introduce the 
single lens approximation to obtain the approximate position 
of the image and make triangles around it ``uncertain'' until the image is picked up. 

\begin{figure}[ht!]
\plotone{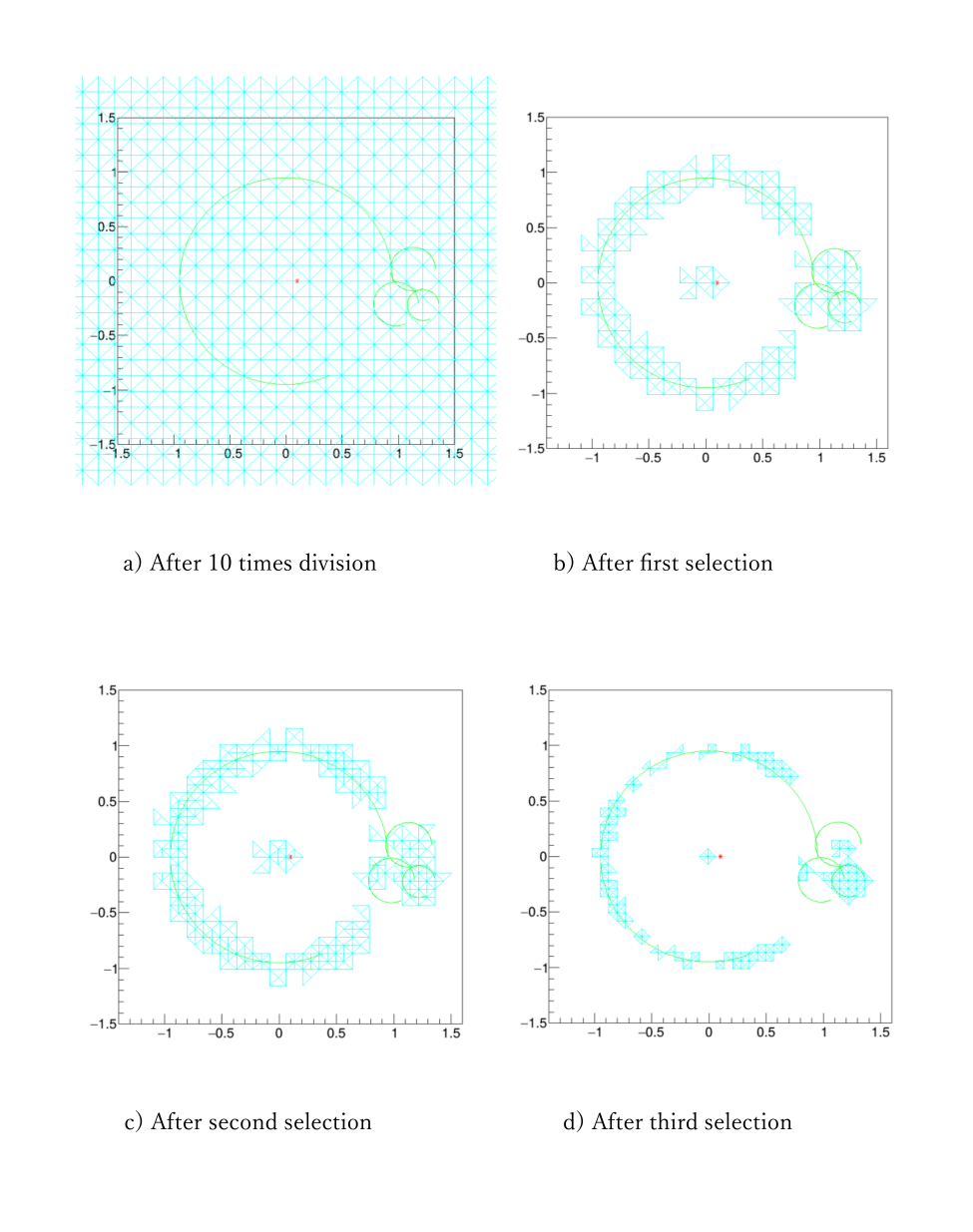}
          \caption{Progress of the selection process. 
          The top left panel (a) shows the triangles on the lens plane before selection. 
          In this stage, there are 2048 triangles on the uncertain queue. 
          Panels b, c, and d show the progress of the selection process. 
          The area of the uncertain triangles decreases with this process,  
          but there is still no in triangle at this stage. 
 \label{fig:progress}}
\end{figure}

Figure \ref{fig:progress} shows the progress of the selection process. 
Figure \ref{fig:progress}a shows the triangles on the lens plane before selection. 
Figures \ref{fig:progress}b, \ref{fig:progress}c, and \ref{fig:progress}d show the 
same plots after the first, second, and third selections, respectively. 
The outside areas are gradually increasing and the uncertain areas are decreasing. 
Repeating these selections, images are found to be ``in'' triangle areas. 

\begin{figure}[ht!]
\plotone{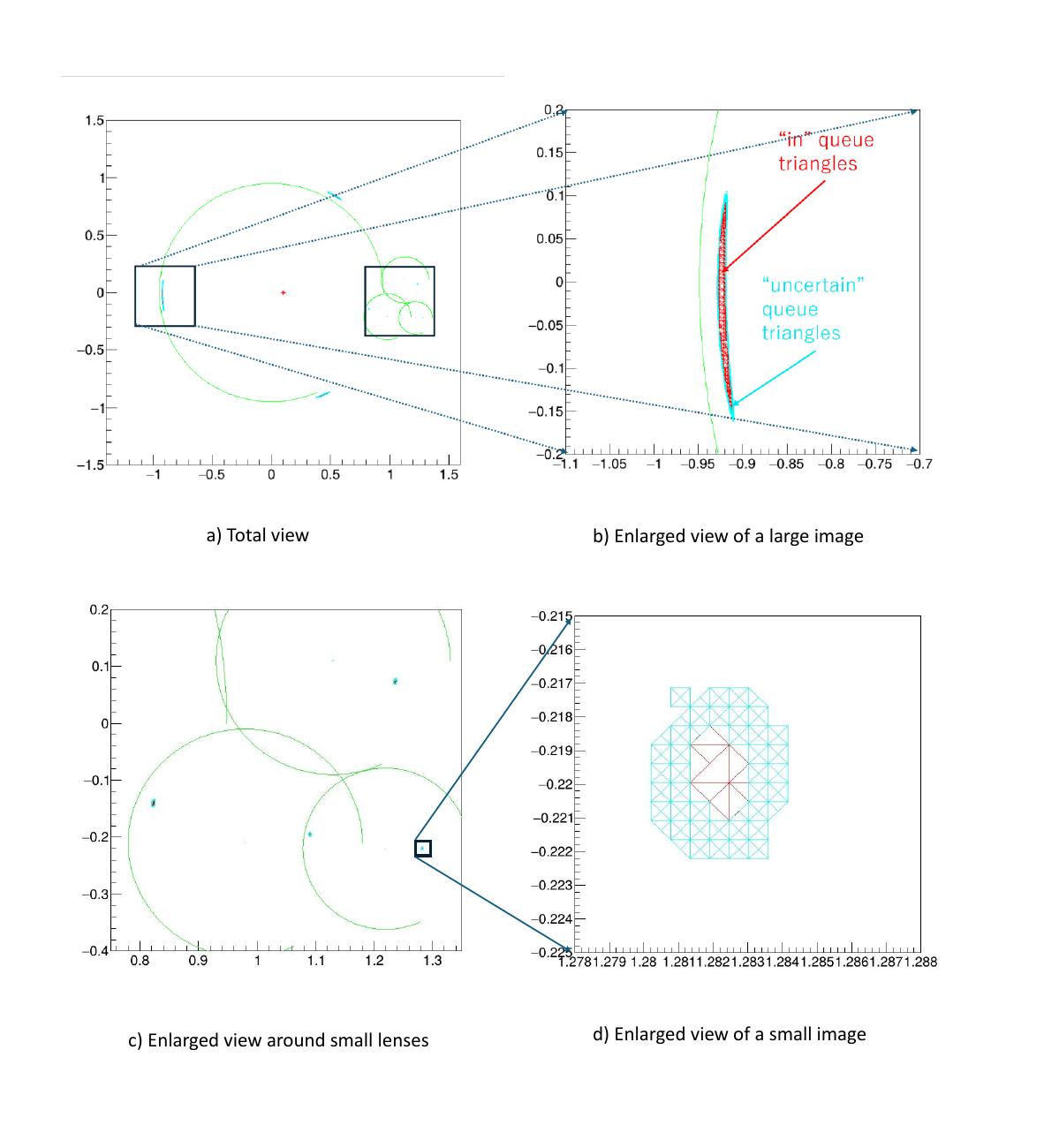}
          \caption{Reconstructed images after 15 times selection.
 \label{fig:result}}
\end{figure}

Figure \ref{fig:result} shows the triangles after 15 selection rounds. 
Here, we assumed $\rho = 0.01$. 
The images were reconstructed using ``in'' and  `uncertain'' triangles. 
Figure \ref{fig:result}a shows the entire lensing area. Figures \ref{fig:result}b, \ref{fig:result}c, 
and \ref{fig:result}d 
show enlarged views of the images. 
The ``uncertain'' triangles surrounding the ``in'' triangles show the border of the images. 
The lens configuration was the same as that of Fig. \ref{fig:contour}, which 
clearly shows all of the images were successfully picked up. 

\begin{figure}[ht!]
\plotone{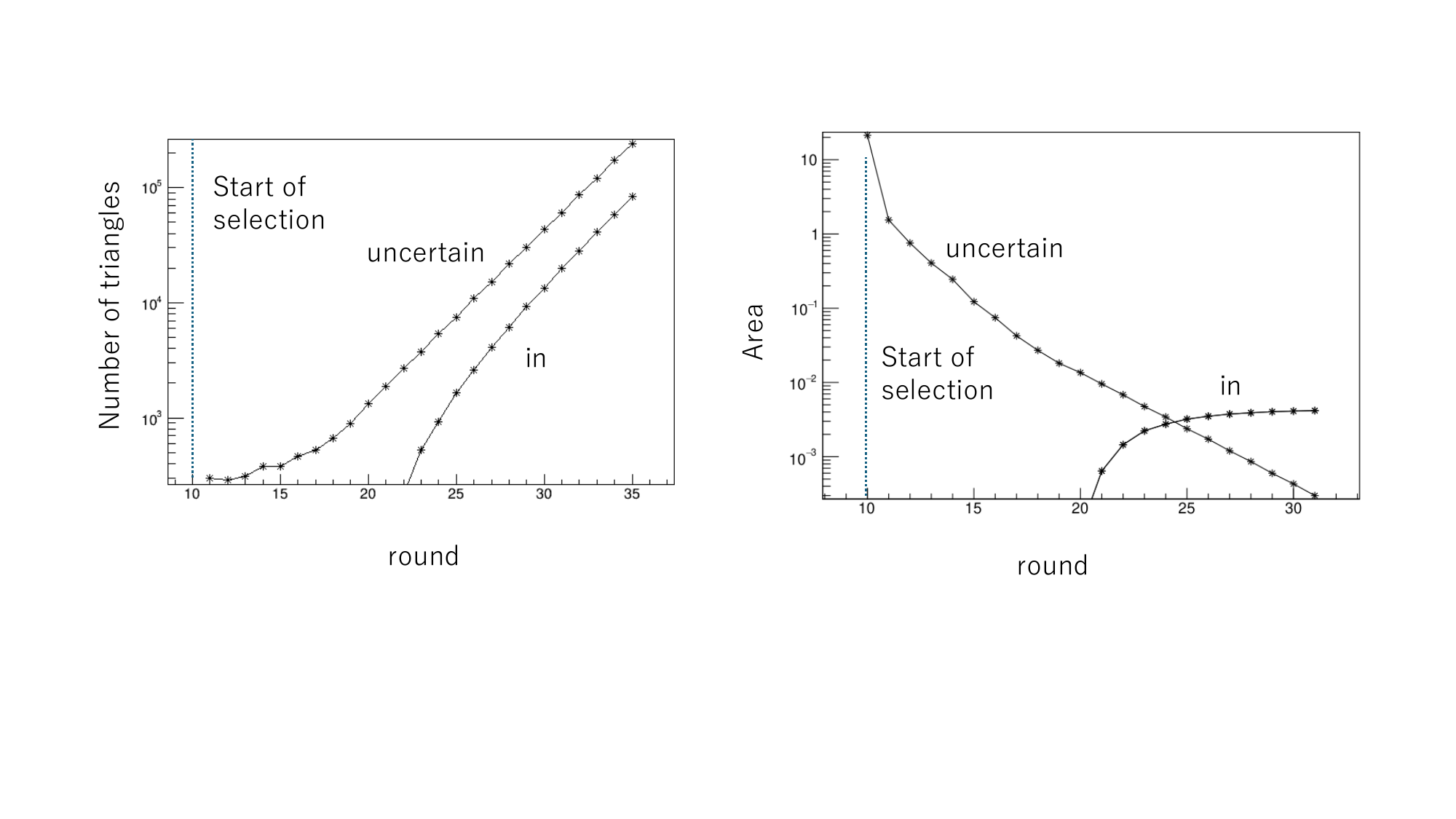}
          \caption{Progress of the total number and total areas of triangles.
 \label{fig:uncertain}}
\end{figure}

The selection process is still useful after finding images to integrate. 
If the surface brightness of the source star is constant, 
magnification $A$ can be obtained based on the ratio of the 
total area of the images to the area of source star $S_s$. 
The total area of the images can be obtained through the  
summation of the areas of the images. 
The summation of the areas of the ``in'' triangles, $\sum_{in} S_i$, gives the lower limit of the 
total area of the images, where $S_i$ is the area of an ``in'' triangle.
The summation of the areas of the ``in'' triangles and  ``uncertain'' triangles, 
$\sum_{in} S_i + \sum_{uncertain} S_i$, gives the upper limit of the total area. 
The selection process should be repeated to reduce uncertainty. 
The uncertainty of the total area is expected to decrease as $2^{-n/2}$,
where $n$ is the number of repetitions of the selection process. 
The selection process must be continued until the uncertainty 
of the magnification becomes sufficiently small. 
Figures \ref{fig:uncertain}a and \ref{fig:uncertain}b show the progress in 
the numbers of triangles 
and their total area, respectively.
The number of ``uncertain'' triangles continues to increase with the processing round. 
The number of ``in'' triangles also increase after finding the first one. 
The total area of ``uncertain'' triangles seems to converge to zero as expected, and the total area 
of ``in'' triangles seems to converge to some value. 
We introduce the first approximation of the total area 
$S_{IU} \equiv \sum_{in} S_{i} + \sum_{uncertain} S_{i} / 2$. 
However $S_{IU}$ does not convergence quickly.

\begin{figure}[ht!]
\plotone{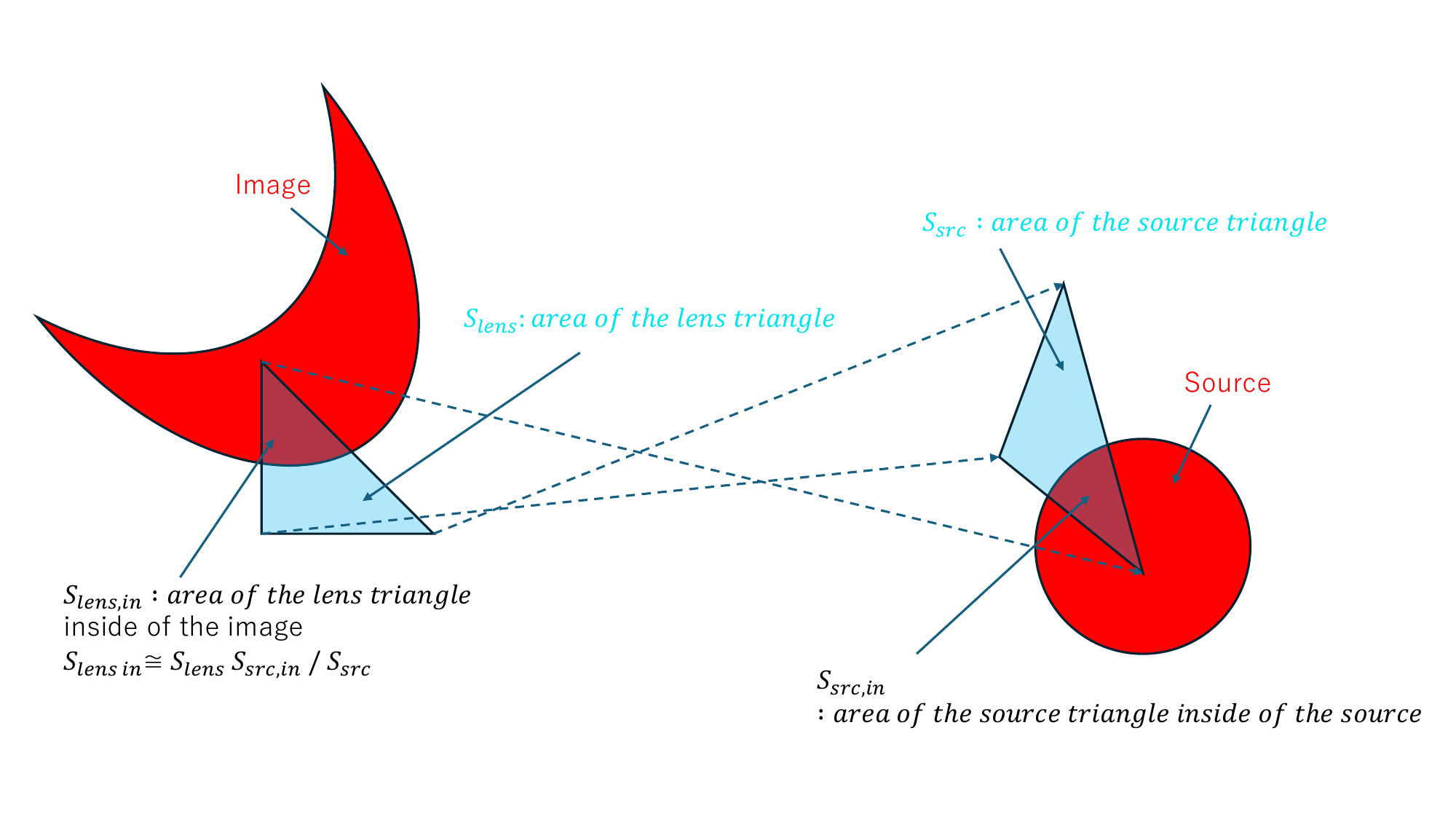}
          \caption{Schematic diagram of the method used to obtain a better approximation 
          of the total area of the images. 
 \label{fig:approximation}}
\end{figure}

\begin{figure}[ht!]
\plotone{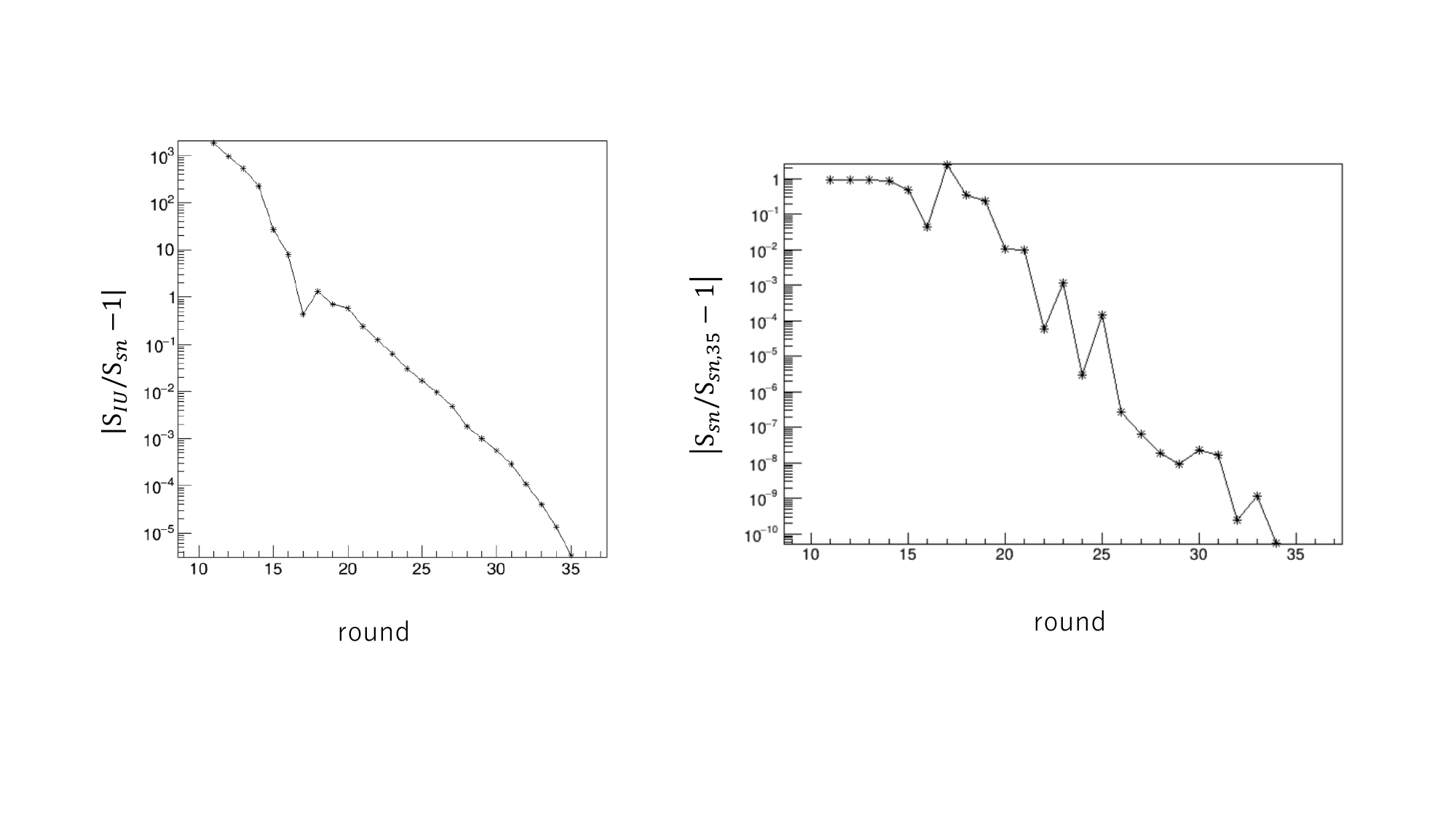}
          \caption{Convergence of the calculated total areas along with the 
          differences between $S_{IU}$ and $S_{sn}$ (panel a) and $S_{sn}$ and $S_{sn,35}$ 
          (panel b) plotted against round. 
 \label{fig:convergence}}
\end{figure}

To improve the convergence speed, we introduced a new approximation. 
Figure \ref{fig:approximation} shows the approximation schematically. 
A portion of the mapped source triangle of an ``uncertain'' triangle 
is inside  the source star 
while the remainder is outside. 
We assumed that the ratios of the inside and outside areas are 
the same for lens and source triangles. 
Then we calculated the ``in'' areas of ``uncertain'' triangles to obtain 
a better approximation 
($S_{sn}$) of the total area of the images. 
Figure \ref{fig:convergence}a shows the difference between $S_{IU}$ and $S_{sn}$, 
They appear to be converging to the same value. 
Figure \ref{fig:convergence}b shows the convergence of the new approximation. 
In this plot, we assumed that the value after 35 rounds ($S_{sn,35}$) was 
the correct value, and 
the differences are plotted. 
The convergence of the new approximation seemed to be much faster that of than that of the old one. 

\begin{figure}[ht!]
\plotone{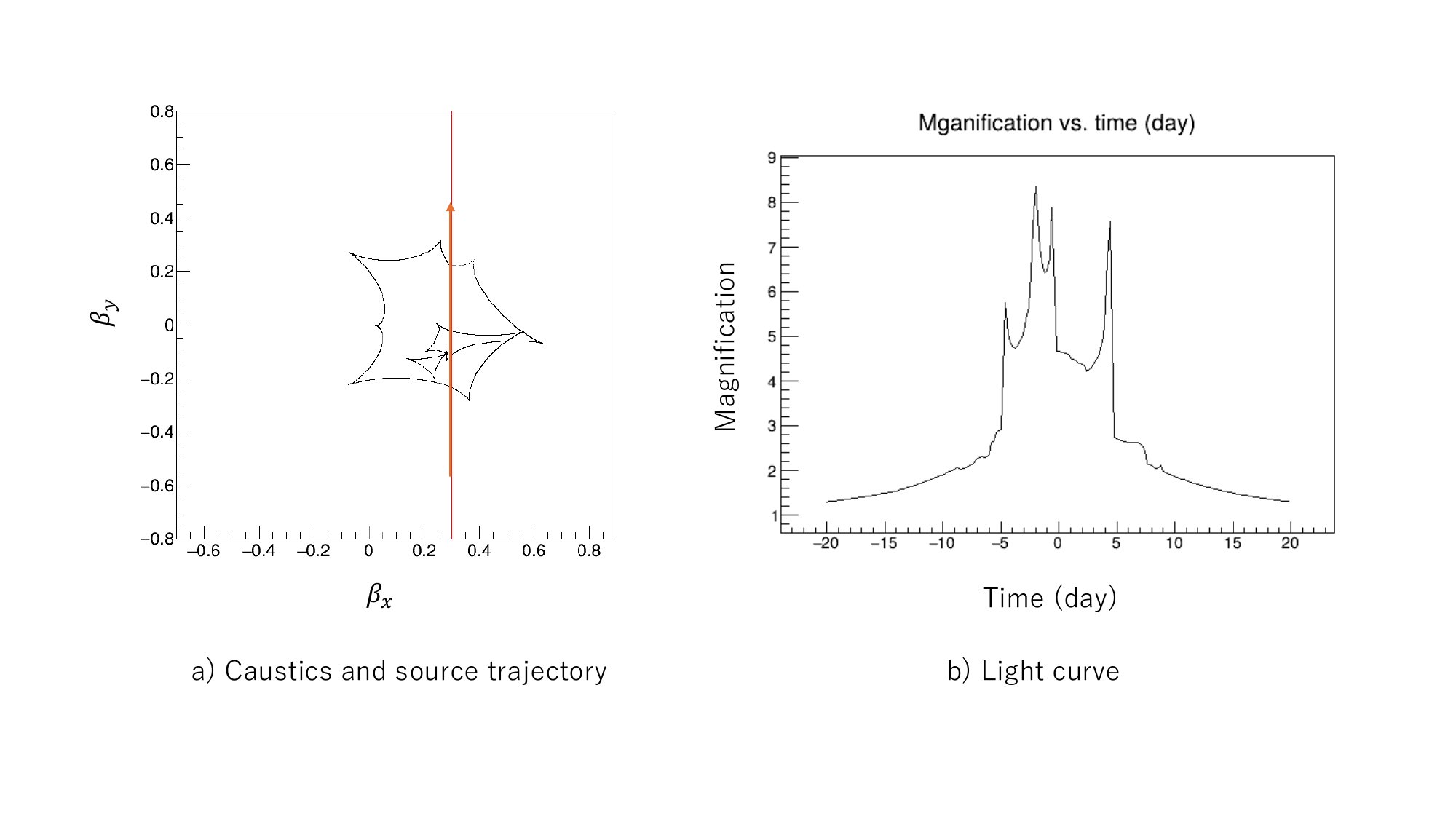}
          \caption{Caustics and source trajectory (a), along with light curve (b), 
          where the source was assumed to pass (0.3, 0.0) at time 0  and move 
          in the y-direction. 
          Einstein radius crossing time $t_E$ was assumed to be 20 days. 
 \label{fig:lightcurve}}
\end{figure}

A light curve was calculated using this algorithm. Figure \ref{fig:lightcurve}a shows 
caustics and the assumed source trajectory. 
The caustics were obtained as mapped critical curves to the source plane. 
The critical curves were obtained as high magnification $|1/J(\vec{\theta})| \gg 1$ points. 
However, calculating $|J(\vec{\theta})|$ with Eq. \ref{eq:J} was found to be very time consuming. 
Thus, we used the ratios of the areas of the lens and source triangles instead. 
Using the assumed source trajectory shown in Fig. \ref{fig:lightcurve}a and $\rho = 0.01$, 
we obtained the light curve shown in Fig. \ref{fig:lightcurve}b. 
Multiple peaks caused by multiple caustic crossings are clearly shown. 

The validity of identifying whether a triangle is inside or outside the source is a critical issue in this algorithm.  
To assess this validity, we employ an integrity check with the following procedure:
\begin{enumerate}
  \item Triangle pairs identified as ``out'' are retained in the ``out'' queue.
  \item All triangle pairs in the ``out'' queue are subdivided into smaller triangles.
  \item Each of the subdivided triangles is then re-evaluated to determine whether it is inside, outside, or uncertain.
\end{enumerate}
If the original identification is valid, all subdivided triangles should also be classified as ``out.''  
We have performed this integrity check for several configurations, and no ``in'' or ``uncertain'' triangles were found among the subdivisions.  
Similar tests conducted on ``in'' triangle pairs also did not yield any ``out'' or ``uncertain'' cases.

In the above discussion, we have assumed that the surface brightness of the source star is uniform. 
However, a non-uniform surface brightness distribution—most notably limb darkening—is 
a well-known effect \citep{1921MNRAS..81..361M}. 
Limb darkening is typically symmetric, with the brightness being nearly constant 
near the center and declining steeply toward the limb.
In our method, the approximate brightness contribution of a triangle can be estimated 
from the surface brightness values at its vertices. 
These values can be obtained from a given limb darkening model, allowing the 
total magnification to include this effect. 
Since our method samples the source plane more densely near the limb, 
it is well suited to handle limb darkening. If further resolution is required, 
additional subdivision of triangles can be applied.
Moreover, irregular surface brightness variations such as stellar spots 
\citep{2000MNRAS.316..665H} can also be accommodated in our framework, provided 
that a surface brightness model is available. In such cases, 
localized refinement of the triangle mesh around the spot region 
would be necessary to accurately capture the photometric features.

\section{Implementation in c++} \label{sec:implementation}
We developed a pilot system for the algorithm. This system was implemented in c++. 
We first used c language to build the pilot system. 
However we found that its implementation in the c language required significant work. 
On the other hand, c++ has sufficient functionality to implement complex algorithms. 
We used class, template, overload, etc to develop the pilot system.
We also used ROOT class libraries (\url{https://root.cern.ch/}) and PyROOT (https://root.cern/manual/python/) 
to create graphics and an interface to Python.
Although the ROOT class library has significant functionality, 
we only used it to create graphics. 
Most of the functions of our pilot system could be used without ROOT. 
The interface to Python with ROOT and PyROOT was very useful. We did not have to make 
a special definition file. All of the classes and functions in our C++ system can be 
obtained from Python. 

The source files were organized as conventional header files (.h) and source files (.cpp). 
The classes and functions were defined in the header and source files. 
There are seven pairs of header and source files : Geom, Lens, Source, Fractal, Motion, Lcurve, and RT-Graph.
The Geom files defined two-dimensional and three-dimensional geometrical classes and functions.
The Lens files defined gravitational lensing classes. 
The Source files defined the Source classes.
The Fractal files defined the classes used in the fractal algorithm. 
The Motion files defined the classes for the motions of the source, lenses, etc.
The Lcurve files defined the light curve classes. 
The RT-Graph files defined the classes used to draw objects. 
All of the classes except for in the RT-Graph files 
were defined for the float (32-bit floating point), double (64-bit floating point), 
and long double (128-bit floating point) types.
In our algorithm, calculations are expected to be almost free from singularities. 
Calculations using the double type are expected to be sufficient for most purposes. 
The float type could be used to upgrade the program for high-performance computing 
using inexpensive graphical processing units (GPUs). 
Recently the inexpensive GPUs used in "gaming PCs" have very high performances 
for the float type, whereas their  
performances for the double type are not very high.
Calculations using the long double type would be overly precise for most purposes. 
However it may be useful for a final check of the calculation. 
In those programs, only RT-Graph uses ROOT class library. Other programs are independent of ROOT.

The current version of the source code is available on \url{https://github.com/fabe758/Frame}.

\section{Discussion} \label{sec:discussion}
As discussed in section \ref{sec:intro}, most of the previous methods 
had a limitation on the number 
of lenses that could be used. 
However, the present method can calculate the lensing effect with no limitation 
on the number of lenses. 
In principle, the calculation time is proportional to the number of lenses. 
Significant CPU power would be needed to fit data because a wide parameter 
space must be searched to find the best parameters. 
Although the problem to search large parameters space is still remained, 
most of the problems for the multiple lens analysis could be overcome. 
The round-off error problem could also be overcome. 
Because we do not attempt to solve the lens equation, 
no caustic singularity appears in the 
calculation. 
To test for round-off errors, we performed the same calculations 
 using the float (32-bit floating point), 
double (64-bit floating point), and long double (128-bit floating points) types 
and compared the results. 
The difference was $0.00122 \%$ between the float and long double types, 
and $4.2296608 \times 10^{-14}$
between the double and long double types. 
Of course, these values depend on the situation. 
However the result for the double type seemed reasonably 
good and useful for most purposes, 
while the float type seemed useful for quickly finding values. 
In addition to the advantages discussed above, the present algorithm is reasonably fast. 
We performed benchmark tests using this algorithm with the float, double, and long double types. 
The light curves for 1000 points were produced in $1.87$, $2.17$, 
and $6.60$s using the float, double, and long double data types, respectively.
Compared to inverse-ray shooting, the CPU times were dramatically reduced. 
Ordinarily, inverse-ray shooting requires at least several hours to obtain a meaningful 
magnification map. 
In this benchmark, we made 15 times selections (round = 25). 
This corresponded to approximately 4 to 5 digit precision for $\rho = 0.01$. 
If a higher precision or smaller $\rho$ is needed, additional 
selections would be necessary. 
Thus, high-precision or the processing of a small $\rho$ needed longer CPU times. 
We used an ASUS laptop PC that had an AMD Ryzen AI 9 HX 370 CPU. 
This is a very powerful machine for a laptop PC. 
However it is not as fast as a typical server. 
We used a Linux distribution EndeavourOS running on Windows Subsystem for Linux (WSL). 
Thus there was some overhead caused by the virtual machine. 
Although the overhead did not seem to be large, 
the effective performance would be lower than that of the bare Linux on the same machine. 
It should also be noted that the present program is single thread, 
and only one core of the 12 cores was used for this calculation. 
In principle, the present program could be rewritten for concurrent processing. 
If 12 cores could be effectively used, the speed would be approximately 10 times faster. 
If GPUs could be effectively used, a speed increase by a factor of 100 or more could be achieved. 

Finding a very small images is a weak point of the present algorithm. 
Some may be lost. 
If the source star is far from the lens system, very small images are produced 
close to the centers of the lenses. 
Because they are very small, most of them are lost. 
However, their contribution to the total magnification is expected to be small 
because they are very small. 
On the other hand, a small source star produces small images. 
In this case, additional divisions and selections are necessary. 
The side length of the lens triangle decreases by a factor of $1/\sqrt{2}$ 
after division. 
Thus, seven additional divisions would be necessary to find images produced 
a source star that was a 10 times smaller. 
To find the equivalent images of source star of $\rho = 0.001$, 22 times selections 
are needed and it took $2.08$, $2.37$, $7.23$s using the float, double, 
and long double data types, respectively.
In the real microlensing analyses, image sizes varies with time, fixed number 
of division would be not appropriate. 
Stopping the divisions after reaching a fixed number of triangle pairs might be 
more appropriate. 

The above discussion focused on the modeling of a multiple lens system 
for a microlensing planet search. 
All lenses were assumed to be on the same plane, 
but this is not essential. 
The present algorithm could easily be expanded to analyze more general problems 
such as quasar microlensing (\citet{1986A&A...166...36K}).
In Eq.\ref{eq:lens}, we assumed that $\sum_i q_i = 1$ and all of the vectors 
($\vec{\beta}, \vec{\theta},$ and $\vec{l_i}$) were normalized by $\theta_E$. 
However, these constraint can be removed. 
Thus, we introduce the following more general lens equation as, 

\begin{equation} \label{eq:general}
          \vec{\beta} = \vec{\theta} - \sum^{n}_{i=1} m_i 
          \frac{D_S - D_{L,i}}{D_S} \frac{\vec{\theta} - \vec{l_i}}{|\vec{\theta} - \vec{l_i}|^2},
\end{equation}

where $m_i$ and $D_{L,i}$ are the mass of the ith lens and distance to the ith lens from the observer, 
respectively. 
This means our algorithm could be applied to quasar microlensing by simply replacing 
$q_i$ with $m_i \frac{D_S - D_{L,i}}{D_S}$.

\section{Summary} \label{sec:summary}
In summary, a new algorithm to analyze a multiple lens system was introduced. 
Because this algorithm does not need to solve the lens equation, there is no limitation on the 
number of lenses and no caustic singularities. 
This algorithm adopts a fractal-like self-similar division method to reduce 
the CPU time. 
Compared to inverse-ray shooting, the CPU time was dramatically reduced. 
The calculation precision and the CPU time have a trade-off relationship. 
We can obtain a quick result by reducing the number of divisions and selections.
Increasing the number of divisions and selections make it possible to obtain more precise values. 
The precision of the calculation using the double type 
is sufficient for most purposes and a calculation using the float type may be useful for a quick result. 
Although the present program is still being developed and may not be 
fast enough to search a large parameter space 
to identify a multiple planet system, further development and a speed increase through parallel processing 
are expected. 

Improving this algorithm and the program will make it possible to explore multiple exoplanet 
system and search for exomoons in future space based microlensing observations. 
Application of the new algorithm to future analyses of quasar microlensing are also expected.

\begin{acknowledgments}
The author would like to thank Professor Valerio Bozza of University of Salerno 
for his valuable advises and comments. 
The author was supported by a grant in aide for scientific research 
          from the Japan Society for the Promotion of Science (JSPS 19KK0082).
\end{acknowledgments}


\bibliography{fractal.bib}{}
\bibliographystyle{aasjournal}



\end{document}